\documentstyle[preprint,floats,prl,aps]{revtex}
\input epsf.tex
\tightenlines

\def\eqn{\begin{equation}}
\def\endeqn{\end{equation}}
\def\eqna{\begin{eqnarray}}
\def\endeqna{\end{eqnarray}}

\begin{document}
\draft
\title
{
\rightline{CLNS98/1599}
\rightline{hep-ph/9812443}
Accuracy of Calculations Involving
$\alpha^3$ Vacuum-Polarization Diagrams: Muonic Hydrogen Lamb Shift
and Muon $g-2$
}

\author{
T. Kinoshita\thanks{e-mail: tk@hepth.cornell.edu}
 }

\address{ Newman Laboratory of Nuclear Studies,
Cornell University, Ithaca, NY 14853 }

\author{
M. Nio\thanks{e-mail: makiko@phys.nara-wu.ac.jp}
 }

\address{ Department of Physics,
Nara Women's University, Nara, Japan 630 }

\date{\today}
\maketitle

\begin{abstract}
The contribution of
the $\alpha^3$ single electron-loop vacuum-polarization diagrams to
the Lamb shift of the muonic hydrogen
has been evaluated recently by two independent methods.
One uses
the exact parametric representation
of the vacuum-polarization function
while the other relies on the Pad\'{e} approximation method.
High precision of these values offers an opportunity
to examine the reliability of the Monte-Carlo integration
as well as that of the Pad\'{e} method.
Our examination covers both
muonic hydrogen atom and  muon $g-2$.
We tested them further for the cases involving
two-loop
vacuum polarization, where an exact analytic result is known.
Our analysis justifies the result for the Lamb shift
of the muonic hydrogen and also
resolves the long-standing discrepancy
between two previous evaluations of the muon $g-2$ value.
\end{abstract}

\vspace{5ex}


\section{Introduction}
\label{sec:intro}


In a recent paper we evaluated the contribution  $\Delta E^{(p6)}$
of the sixth-order electron-loop
vacuum-polarization diagrams to the muonic hydrogen $(\mu^- p^+)$ Lamb Shift 
\cite{ourpaper}. 
Together with the proposed measurement of the $2P_{1/2}-2S_{1/2}$
Lamb Shift \cite{taqqu} it will lead to a very precise determination
of the proton charge radius. 

The most laborious part of this calculation is that of
the single electron loop diagrams 
contributing to the sixth-order vacuum-polarization function 
(see Fig. 1). 
Using the parametric representation for this function 
\cite{KL2}  
we find its contribution to the Lamb shift to be
\cite{ourpaper} 
\eqn
\Delta E^{(p6)}= 
0.017~410~(9)~
m_r (Z\alpha)^2 \biggl ({\alpha \over \pi}\biggr )^3,
\label{exact}
\endeqn
where $m_r$ is the reduced mass of the muon-proton system
and $Z=1$ is the proton charge in units of the electron charge $|e|$. 
As a cross-check, we have also computed 
$\Delta E^{(p6)}$ using the Pad\'{e}-approximation of 
the single electron loop sixth-order vacuum-polarization function
$\Pi_3^{[1]}(z),~z=q^2/4m_e^2$, where $m_e$ is the electron mass 
 \cite{BB}.
Inserting the real part of $\Pi_3^{[1]}$ derived from
Eq. (7) of Ref. \cite{BB}
into Eq. (3) of Ref. \cite{ourpaper}, 
we obtained the Lamb Shift contribution
in the [3/2] and [2/3] Pad\'{e} approximations.
Since they are practically indistinguishable,
we will not label them separately and simply quote them as
\eqn
\Delta E^{(p6)}_{Re}=0.017~414~9~(25)~
m_r (Z\alpha)^2 \biggl ({\alpha \over \pi}\biggr )^3~.
\label{ReBB}
\endeqn
We also derived the imaginary part of the approximate $\Pi_3^{[1]}$ and, 
inserting it in Eq. (6) of Ref. \cite{ourpaper},
obtained
\eqn
\Delta E^{(p6)}_{Im}=0.017~414~9~(26)~
m_r (Z\alpha)^2 \biggl ({\alpha \over \pi}\biggr )^3~.
\label{ImBB}
\endeqn
Again the [3/2] and [2/3] Pad\'{e} approximations
are nearly indistinguishable.
The results (\ref{ReBB}) and (\ref{ImBB}) are consistent with each other
and agree within one standard deviation
with the direct calculation (\ref{exact}).
Note, however, that the uncertainties in (\ref{ReBB}) and (\ref{ImBB})
are those caused by numerical treatment of the Pad\'{e}
approximation and do not represent the accuracy
of the Pad\'{e} method itself.
To gain insight in how good the Pad\'{e} approximation is
we have examined two cases where exact results are known.
Based on these results
we argue that
the true value of the muonic hydrogen Lamb shift
will be found within 0.000 7 percent of the
Pad\'{e} value, which is well within the uncertainties
quoted in (\ref{ReBB}) or (\ref{ImBB}).

\begin{figure}
\centerline{\epsfbox{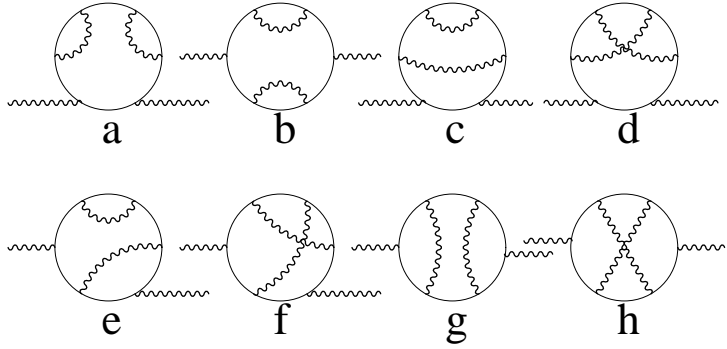}}
\caption{Sixth-order vacuum-polarization diagrams
with a single electron loop.}
\label{figlsp6}
\end{figure}


\section{Derivation of Pad\'{e} approximation results}
\label{sec:outline}


Let us begin with a brief review of the derivation 
of the Pad\'{e} approximation to the 
sixth-order vacuum-polarization function 
$\Pi_3^{[1]} (z),~ z \equiv q^2/4m^2$, by Baikov and Broadhurst
\cite{BB}. 
The analytic properties of $\Pi_3^{[1]} (z)$ they utilized are:
(a) The first three coefficients of the Taylor expansion around $z= 0$.
(b) The first two $\log z$-dependent 
coefficients of the expansion in $1/z$ for large negative $z$.
(c) The threshold Coulomb behavior which is determined by
nonrelativistic quantum mechanics.
Taking account of these informations
they constructed the function 
\eqn
 \tilde{\Pi}_3^{[1]}(z) \equiv
\Pi_3^{[1]} (z) +  4 \Pi_2 (z) + (1-z)G(z)\biggl( {9\over4} G(z) +
{31 \over 16} + {229 \over 32z} \biggr ) - { 229 \over 32z} 
- {173 \over 96} ,
\label{pi3}
\endeqn
where
$\Pi_2(z)$ is the fourth-order vacuum-polarization function \cite{kallen}.
$G(z)$ is the hypergeometric function $ _2\!F_1(1,1;{3 \over2};z)$.
For the negative real $z$, $G$ is given by 
\eqn
G(z)={ 1 \over \sqrt{z^2-z} } \ln( \sqrt{-z}+\sqrt{1-z} ).
\endeqn
Analytic continuation of $G(z)$ 
from $z < 0$ to $ z \ge 1$ through the upper $z$-plane yields
\eqna
&& {\rm Re} G(z)={- 1 \over \sqrt{z^2-z} } \ln( \sqrt{z}+\sqrt{z-1} ),
\\
&&{ \rm Im} G(z)={\pi \over 2 \sqrt{z^2-z} } ~. 
\endeqna

By construction
$\tilde{\Pi}_3^{[1]}(z)$ 
is analytic in the $z$-plane cut along the real axis $1 < z < \infty$.
The function defined by
\eqn
{1-\omega \over (1+\omega)^2} 
\biggl( \tilde{\Pi}_3^{[1]} -\tilde{\Pi}_3^{[1]}(-\infty)\biggr ), 
~~~z= {4 \omega \over (1+\omega)^2 }~,
\label{deffunc}
\endeqn 
is analytic for $|\omega| < 1$,
and may be simulated accurately by
a Pad\'{e} approximant $P(\omega)$. 
Analytic informations on $\Pi_3^{[1]}$ listed above
are translated into six data for $P(\omega)$:
$\{P(-1)$, $P(0)$, $P'(0)$, $P''(0)$, $P'''(0)$, $P(1)\}$. 
Using these data we have constructed 
\eqn
P(\omega)= { a_0 + a_1 \omega + a_2 \omega ^2 + a_3 \omega^3
\over b_0 + b_1 \omega + b_2 \omega^2 + b_3 \omega^3 }. 
\label{Pomega}
\endeqn
The coefficients $a$'s and $b$'s for both [2/3] and [3/2]
Pad\'{e} approximations calculated from 
the given data are listed in Table I. 
Once a Pad\'{e} approximant is constructed, 
we can readily obtain the corresponding 
$\Pi_3^{[1]}$ from (\ref{pi3}) and (\ref{deffunc}).

\begin{table}
\caption{Coefficients of Pad\'{e} approximants. 
We set $b_0=1$ for the overall normalization. 
\label{table1}
}
\[
\begin{array}{ccc}
\hline \hline
 {\rm Coefficient}    &   {\rm [2/3] Pad\acute{e}}   &   {\rm  [3/2] Pad\acute{e}} 
\\ \hline
a_0 &  ~5.450~103~092~   &  ~5.450~103~092~ 
\\ \hline
a_1 & -0.966~458~776~    &  -0.891~171~812~ 
\\ \hline
a_2 & -1.785~150~929~    &  -1.800~086~980~
\\ \hline
a_3 &  0                 &  -0.025~240~917~
\\ \hline
b_0 &  1                 &   1 
\\ \hline
b_1 &  -0.456~709~419~  &  -0.442~895~559~
\\ \hline
b_2 &  -0.121~731~656~  &  -0.128~331~492~
\\ \hline
b_3 & ~0.001~706~927~    &   0 
\\
\hline \hline
\end{array}
\]
\end{table}

The $2P_{1/2}-2S_{1/2}$ Lamb shift  of the muonic hydrogen
has been evaluated in two ways.
One uses the formula which contains $\Pi (z)$ for negative $z$ \cite{ourpaper}: 
\eqn
\Delta E = 
 { 2 \over \pi} (Z\alpha)^2 m_r
           \int_0^{\infty} da \tilde{\rho}(a^2)
        \Pi(-a^2/(4\beta^2)),
\label{lsRe}
\endeqn
where 
\eqn
\tilde{\rho}(a^2)=  {2 a^2 (1-a^2) \over (1+a^2)^4 }~,
\endeqn
and
\eqn
\beta = { m_e \over m_r Z\alpha} = 0.737~383 ~76~(30)~.
\endeqn
Substituting  $\Pi_3^{[1]}(z)$ determined from
Table I in Eq. (\ref{lsRe}) and evaluating it numerically
\footnote{Unless specified otherwise integrals are
evaluated numerically on DEC$\alpha$ 
using the adaptive-iterative Monte-Carlo
subroutine VEGAS \cite{lepage}.},
we obtained the result (\ref{ReBB}).
  
The second approach utilizes the
imaginary part of the approximate $\Pi_3^{[1]}$ for $z > 1$
obtained from the Pad\'{e} approximant $P(\omega)$ 
by taking
its value on the unit circle  in 
the upper-half $\omega$ plane. 
Using this information
the $2P_{1/2}-2S_{1/2}$ Lamb shift of the muonic
hydrogen can be expressed by \cite{ourpaper}
\eqn
\Delta E = m_r (Z\alpha)^2 \int_1^{\infty}  dz
{{\rm Im} \Pi(z) \over \pi} { 2 \beta^2  \over  (1 + 2\beta \sqrt{z})^4} ~.
\label{lsIm}
\endeqn
The result (\ref{ImBB}) follows from (\ref{lsIm}). 


\section{Discussion}
\label{sec:discussion}


The results (\ref{ReBB}) and (\ref{ImBB})
obtained by the Pad\'{e} approximation method are in good
agreement with the direct result (\ref{exact}).
The difference between them is within one standard deviation
of the result (\ref{exact})
and can be ignored for the purpose of
comparison with experiment.
However, (\ref{exact}),  (\ref{ReBB}) and (\ref{ImBB}) all involve some
uncertainties inherent to their derivation.
Thus it will be worthwhile to examine the nature of
these uncertainties.
There are at least two possible causes
which may contribute to these uncertainties:
One is that the error estimate generated by VEGAS in the
evaluation of (\ref{exact})
might be a gross underestimate of the true error.
The other arises from the 
fact that the Pad\'{e} approximant does not represent
$\Pi_3^{[1]} (z)$ accurately for all values of $z$.


\subsection{Non-statistical error in VEGAS calculation}
\label{subsec:VEGASerror}


The integration routine VEGAS is an adaptive-iterative
procedure based on random sampling of the integrand \cite{lepage}.
In the $i$-th iteration, the integral is evaluated by sampling
the integrands at points chosen randomly according to a
distribution function chosen in the $(i-1)$-st iteration.
This generates the approximate value $I_i$ of the integral,
an estimate of its uncertainty $\sigma_i$, and the distribution
function to be used for the next iteration.
After several iterations $I_i$ and $\sigma_i$ 
are combined under the assumption that all iterations 
are statistically independent.
The combined value and error are given by
\eqna
I &=& (\sum_i ( I_i /\sigma_i^2 ))/(\sum_i (1/ \sigma_i^2 )), \nonumber  \\
\sigma &=& (\sum_i (1/ \sigma_i^2 )) ^{-1/2} .
\label{combines}
\endeqna

In general VEGAS is found to converge rapidly 
for a sufficiently large number of samplings.
However, a special care is required when it is applied to
the integration of Feynman amplitudes in which 
renormalization is carried out on the computer
relying on point-by-point cancellation of singularities
between the unrenormalized integrand and the corresponding
renormalization term.
This does not pose a problem if we operate with infinite precision.
In reality, however, calculation is carried out using finite
precision arithmetic, such as double (real*8)
or quadruple (real*16) precision. Random numbers generated
by VEGAS for sampling of the integrand will inevitably hit
points very close to some singularity.
This will result in evaluation of the difference of two
very large and nearly identical 
 numbers with finite amount of significant digits.  
At such a point most of the significant digits
cancel out leaving
only a few significant digits 
or no significant digit at all.
(This will be referred to as digit-deficiency or $d$-$d$ problem,
and the subdomain of integration where this happens,
which contains some boundary surface of the hypercube in our problem,
will be called the $d$-$d$ domain.)
This introduces nonstatistical noise in the evaluation of the integral
and its error estimate, 
even though the effect,
being confined to the $d$-$d$ domain of very small measure,
 is often not
readily distinguishable from the fluctuations
inherent to random sampling of the integrand.
They tend to give
the integral a false value and might cause 
deviation of error estimate 
from the assumed statistical behavior.
However, the $d$-$d$ problem encountered in the $i$-th iteration
will not affect the performance of the $(i+1)$-st iteration
unless it distorts the distribution function very severely.
As is seen from (\ref{combines})
$I_i$ with larger $\sigma_i$ are given smaller weights,
giving only small impact on the composite $I$ and $\sigma$.
Problem arises, however,
if some $\sigma_i$ is relatively small even if it
is suffering from the $d$-$d$ problem.
In such a case the final $I$ may be distorted in an unpredictable way.

Relative impact of the $d$-$d$ domain decreases
as the sampling statistics in an iteration increases.
More importantly, it decreases dramatically
if quadruple precision is adopted.
In the past, however, this was not necessarily a practical
approach because it typically requires
 20 or more computing time for execution
compared with double precision calculation.
Only recently the availability of faster computers 
has made this a viable option.

Because of higher speed we normally
start evaluation of Feynman integrals
in double precision.
If this runs into a $d$-$d$ problem,
we split the integration domain into a small region around
the $d$-$d$ domain and the remainder.
The difficult region is then evaluated in quadruple precision,
while evaluation of the rest continues in double precision.
(In some really difficult cases, it will be preferable
to adopt quadruple precision
for the entire integration domain.)
This strategy has been very successful and many integrals
were evaluated very precisely in this manner.
In fact, in some cases, the achieved numerical precision was
such that it led to uncovering of errors in some analytic
or semi-analytic calculations \cite{tk2}.

Unfortunately, it is not always easy to detect problems
caused by the $d$-$d$ distortion.
This was the case with some muon $g-2$ diagrams involving
sixth-order vacuum-polarization diagrams \cite{tk1}.
While Ref. \cite{tk1} reported the value
$-0.2415 (19) (\alpha /\pi )^4$
the same diagrams evaluated using
the Pad\'{e} approximation method gave
\cite{BB} 
\eqn
 a_\mu^{(8)} (Pad\acute{e}) = -0.230362~(5) \left ( {\alpha \over \pi} \right)^4 
\label{amu8BB}
\endeqn
which disagreed with Ref. \cite{tk1}
by about 6 standard deviations.
The reliability of the error estimate in \cite{tk1}
was thus called into question.

As was discussed above, the most
effective way to separate the $d$-$d$ error
evaluated in double precision
from the statistical uncertainty 
is to go over to quadruple precision.
We have therefore repeated
the integration of Ref. \cite{tk1} entirely
in quadruple precision with a roughly equal amount of 
sampling statistics.
The new value
\eqn
 a_\mu^{(8)} (Fig. 1) = -0.2285~(18) \left ( {\alpha \over \pi} \right)^4 
\label{amu8}
\endeqn
agrees within one standard deviation with the Pad\'{e} value 
(\ref{amu8BB}) but
disagrees strongly with the old value.

A closer examination of Ref. \cite{tk1} reveals that
some of the eight integrals 
(in particular, the diagram $e$ of Fig. 1, evaluated
with 60 million sampling points per iteration)
show signs of suffering from
the presence of $d$-$d$ error. 
This is now clearly confirmed by the new result (\ref{amu8}).
One way to overcome this problem within double precision calculation
 is to go over to 
much larger statistics.  
In order to see whether
ten times more sampling points per iteration 
is sufficient for the diagram $e$,
we have evaluated it in both double
\footnote{Evaluated on the Fujitsu VX computer at the Computer Center of
Nara Women's University. }
and quadruple\footnote{Evaluated on the IBM SP2 computer
at Cornell Theory Center.}
precisions (with 600 million sampling point per
iteration and 60 iterations).
The two results are in good agreement 
showing that the double precision calculation is reliable now.
Encouraged by this, we have evaluated
the remaining seven integrals in double precision
with ten times more statistics.
The total contribution is found to be
\eqn
 a_\mu^{(8)} (Fig. 1) = -0.230~596~(416) \left ( {\alpha \over \pi} \right)^4  .
\label{newamu8}
\endeqn
This is in good agreement with the Pad\'{e} result (\ref{amu8BB}).

We conclude that the problem encountered by the result \cite{tk1}
was caused solely by insufficient statistics.
Unfortunately this does not become visible until
more extensive calculation is done.

To examine the sensitivity of the Lamb shift calculation
to the precision of the arithmetic used we 
evaluated the sixth-order
vacuum-polarization contribution to the muonic hydrogen
in both double and quadruple precision.           
The results obtained in double and quadruple 
precisions are listed in Table I of Ref. \cite{ourpaper}.
The double precision calculation was carried out
using 100 million sampling points per iteration
while that of quadruple precision was obtained for
1 million sampling points per iteration 
(except for the diagrams $a$ and $e$ which employ 2 and 4 times 
more sampling points, respectively).
As is clearly seen from Table I of Ref. \cite{ourpaper}, 
the results in double and quadruple precision
are in good agreement and give us confidence that
this problem does not suffer visibly from
the $d$-$d$ problem.
The reason why the Lamb shift calculation
looks less susceptible to the $d$-$d$ problem is that, 
unlike the muon $g-2$ case,
the contribution
from large momentum transfer region is strongly suppressed
by the hydrogenic wave function.


\subsection{How good is the Pad\'{e} approximation ?}
\label{subsec:padeapprox}


We now want to examine whether or not
the Pad\'{e} result $a_\mu (Pad\acute{e})$ of (\ref{amu8BB})
agrees within its much smaller error bars  
with the true value of $a_\mu^{(8)} (Fig. 1)$.
This question is raised
because the [2/3] or [3/2] Pad\'{e} approximant $P(\omega)$ of (\ref{Pomega})
does not have complete information on $\Pi_3^{[1]}$.
In principle it is possible to answer this question
by improving the numerical precision of (\ref{newamu8})
by two orders of magnitude.
This is not very practical, however, since it would require
$10^4$-fold increase in the computational effort.
Another approach is to go to higher-order
Pad\'{e} approximations, taking additional
exact properties of $\Pi_3^{[1]}$ into account \cite{chet1,chet2}.

Let us instead follow an alternative and easier route,
namely, examine how well the Pad\'{e} method works
by applying it to the cases where
exact results are known.

The first example is the contribution of
the two-loop vacuum-polarization to the muon $g-2$.
The exact result is given by \cite{laporta}
\eqn
 a_\mu^{(6)} (\alpha^2 v.p.) = 1.493~671~80~(4) 
\left ( {\alpha \over \pi} \right)^3 ,
\label{amu6exact}
\endeqn
where the uncertainty comes only from the uncertainty in the measured value
of $m_\mu /m_e$.
The corresponding value obtained by numerical integration
by VEGAS starting from the K\"{a}llen-Sabry
spectral function is 
\eqn
 a_\mu^{(6)} ({\rm K-S}) = 1.493~672~7~(40) 
                \left ( {\alpha \over \pi} \right)^3 .
\label{amu6}
\endeqn
The difference between (\ref{amu6exact}) and (\ref{amu6})
is well within the error bars of (\ref{amu6}) and
will disappear as the numerical precision of
(\ref{amu6}) increases.

Meanwhile, 
the real and imaginary parts of [3/2] and [2/3] Pad\'{e} approximation 
given in Appendix A produce 
\eqn
 a_\mu^{(6)} ({\rm Pad\acute{e} Re}) = \left \{ \begin{array}{c}
                                1.493~677~6~(16) \\
                                1.493~678~1~(16) \\
                                1.493~676~2~(16) \\
                                1.493~675~9~(16) 
                                           \end{array}
                                 \right \} 
\left ( {\alpha \over \pi} \right)^3 ~,~~~~ 
                                   \begin{array}{c}
                                    \cdots {\rm [2/3]A  } \\ 
                                    \cdots {\rm [2/3]B  } \\  
                                    \cdots {\rm [3/2]A  } \\ 
                                    \cdots {\rm [3/2]B  }  
                                   \end{array}~
\label{amu6Pade}
\endeqn
and
\eqn
 a_\mu^{(6)} ({\rm Pad\acute{e} Im}) = \left \{ \begin{array}{c}
                                1.493~674~2~(38) \\
                                1.493~847~4~(38) \\ 
                                1.493~737~9~(38) \\ 
                                1.493~676~5~(38)  
                                           \end{array}
                                 \right \} 
\left ( {\alpha \over \pi} \right)^3 ~,~~~~ 
                                   \begin{array}{c}
                                    \cdots {\rm [2/3]A  } \\ 
                                    \cdots {\rm [2/3]B  } \\ 
                                    \cdots {\rm [3/2]A  } \\ 
                                    \cdots {\rm [3/2]B  }  
                                   \end{array}  ~
\label{amu6PadeIm}
\endeqn
respectively, where $A$ and $B$ refer to two different
sets of determinations of Pad\'{e} functions.  
%
%
Within the context of this approximation 
it is not possible to tell which of the results (\ref{amu6Pade})
and (\ref{amu6PadeIm}) are good or bad.
Since the $[m/n]$ Pad\'{e} approximation will get closer
and closer to the exact function as $m$ and $n$ increase,
however,
the Pad\'{e} results based on 
the real and imaginary parts should give identical results
in the large $m$ and $n$ limit. 
This means that near-equality of results from the real and imaginary
parts may be chosen as a working criterion for selecting
good Pad\'{e} approximation.
According to this criterion 
the results [2/3]A and [3/2]B are good while
[2/3]B and [3/2]A are not.
Since this is not a rigorous criterion,
we may of course choose the worst case as a measure
of the uncertainty of the Pad\'{e} approximation.


The disagreement between better values of (\ref{amu6Pade}) 
and (\ref{amu6PadeIm})
and the exact value (\ref{amu6exact})  
is about 0.000006
$(\alpha/\pi )^3$  ($\sim$  0.0004 percent), which
falls somewhat outside the estimated errors of (\ref{amu6Pade})
and (\ref{amu6PadeIm}).
The difference 
of about 0.00017
$(\alpha/\pi )^3$ ($\sim$  0.01 percent)
between (\ref{amu6exact}) and 
the $``$worst" value of (\ref{amu6Pade}) 
and (\ref{amu6PadeIm}) is much larger,
and is well outside its estimated error.
Based on these observations we may conclude that
the near-equality of real and imaginary Pad\'{e} values
is a useful working criterion and indicates
that the [2/3] and [3/2] Pad\'{e} methods give
results, with a high probability,
 within 0.0004 percent of the exact value.
Analogously, the Pad\'{e} approximation value (\ref{amu8BB}) 
of an eighth-order muon $ g-2$ may not deviate from the true value 
by more than 0.001 percent,
even allowing for possible differences between
$\Pi_2$ and $\Pi_3^{[1]}$,
 staying within the error bars of (\ref{amu8BB}).  

The second example is
the Lamb shift caused by the fourth-order vacuum-polarization function. 
The Lamb shift evaluated by VEGAS using 
the exact vacuum-polarization function 
\cite{kallen} is
\eqna
\Delta E^{(p4)}_{exact}&=&0.045~922~738~(57)~
m_r (Z\alpha)^2 \biggl ({\alpha \over \pi}\biggr )^2~,
\label{deltaEp4}
\endeqna
while the real  and imaginary parts of the [3/2] and [2/3] Pad\'{e} approximation 
described in Appendix A give
\eqn
\Delta E^{(p4)}_{PadeRe}=
                      \left \{          \begin{array}{c}
                                0.045~923~200~88 \\
                                0.045~923~320~16 \\ 
                                0.045~922~988~50 \\
                                0.045~922~938~13 
                                           \end{array}
                      \right \} 
m_r (Z\alpha)^2 \biggl ({\alpha \over \pi}\biggr )^2~,
                                   \begin{array}{c}
                                    \cdots {\rm [2/3]A  } \\ 
                                    \cdots {\rm [2/3]B  } \\ 
                                    \cdots {\rm [3/2]A  } \\ 
                                    \cdots {\rm [3/2]B  }  
                                   \end{array}~
\label{deltaEp4PadeRe}
\endeqn
and 
\eqn
\Delta E^{(p4)}_{PadeIm}=
                      \left \{          \begin{array}{c}
                                0.045~923~200~97 \\
                                0.045~933~403~92 \\ 
                                0.045~926~642~63 \\
                                0.045~922~938~22 
                                           \end{array}
                      \right \} 
m_r (Z\alpha)^2 \biggl ({\alpha \over \pi}\biggr )^2~,
                                   \begin{array}{c}
                                    \cdots {\rm [2/3]A  } \\ 
                                    \cdots {\rm [2/3]B  } \\ 
                                    \cdots {\rm [3/2]A  } \\ 
                                    \cdots {\rm [3/2]B  }  
                                   \end{array}~
\label{deltaEp4PadeIm}
\endeqn
respectively\footnote{The results (\ref{deltaEp4PadeRe}) and (\ref{deltaEp4PadeIm}) 
are obtained using the numerical integration routine of Maple V. }.
Uncertainties of all numerical coefficients 
in (\ref{deltaEp4PadeRe})
and (\ref{deltaEp4PadeIm})
do not exceed $0.5 \times 10^{-9}$.


The [2/3]A and [3/2]B results of 
(\ref{deltaEp4PadeRe}) and (\ref{deltaEp4PadeIm}) 
are in good agreement with (\ref{deltaEp4}),
the difference being about  0.0007 percent.
Meanwhile the [2/3]B  value of (\ref{deltaEp4PadeIm}) deviates 
from (\ref{deltaEp4}) by about 0.02 percent.
The six standard deviation difference
between (\ref{deltaEp4}) and $``$good" values of
(\ref{deltaEp4PadeRe}) and (\ref{deltaEp4PadeIm}) 
is unlikely to be attributable to the fault of numerical integration.
Instead it must be interpreted as the 
measure of the approximate nature of the [2/3] and [3/2] Pad\'{e},
which can be remedied only by going to higher-oder
Pad\'{e} approximation.

Finally one may ask how can one justify
the use of the Pad\'{e} approximation
for the two-loop vacuum-polarization function
as a model for the three-loop case,
in view of the fact that the three-loop function $\Pi_3^{[1]}$
 has a threshold
singularity ($\sim 1/v$) while $\Pi_2$ does not.
Our answer is that
there is not much analytical difference between $P(\omega )$ of
two-loop and three-loop cases
since this singularity is 
removed explicitly in the construction (\ref{deffunc}) of
the Pad\'{e} function $P(\omega )$.


\subsection{Conclusion}
\label{subsec:concl}


Based on these considerations 
we conclude:

(I)  
The Lamb shift calculation by 
numerical integration method gives a reliable result 
even if double precision arithmetic is used,
provided sufficient sampling of the integrand
is made in each step of iteration.

(II) 
The [2/3] and [3/2] Pad\'{e} methods for the vacuum-polarization function
given in \cite{BB} is a very good approximation
over nearly all momentum space region, 
even though it is exact only at a few chosen 
values of the momentum.
It will be possible to narrow the gap between Pad\'{e} and exact
results by using more input date and by 
going to higher-order Pad\'{e} approximations.
However, we believe that [2/3] and [3/2] cases are good enough for
the purpose of confirming the reliability
of numerical integration method.

\acknowledgments

We thank K. G. Chetyrkin, J. H. K\"{u}hn, R.Harlander,
and M. Steinhauser for bringing our attention to
Refs.\cite{chet1,chet2}.
The work of T. K. is supported in part by the U. S. National Science Foundation.
The work of M. N. is supported in part by the Grant-in-Aid (No. 10740123)
of the Ministry of Education, Science, and Culture, Japan.  
Part of T. K.'s work is carried out on the SP2 computer
at the Cornell Theory Center.


\appendix

\section{Pad\'{e} approximation to $\Pi_2$ }
\label{appendixPade}


The proper two-loop vacuum-polarization function $\Pi_2$
can be obtained from  Eq. (57) of Ref. \cite{kallen}
by removing the contribution of $(\Pi_1 )^2$, 
where $\Pi_1$ is the one-loop vacuum-polarization function.
Since $\Pi_2$ is known explicitly, it can be used to 
construct a Pad\'{e} approximant of any degree of precision.
Our purpose here is to test the approximation to $\Pi_3^{[1]}$.
Thus we construct a Pad\'{e} approximant of $\Pi_2$ 
using analytic information similar to those
used in constructing $\Pi_3^{[1]}$ described in Sec. 
\ref{sec:outline}.
They are:  

\noindent
(i) Small momentum behavior of 
$\Pi_2 (z)$
given by \cite{BB}
\eqn
\Pi_2 (z) = {82 \over 81} z + {449 \over 675} z^2 + {249916 \over 496125} z^3 
+ \cdots ~.
\label{pi21smallz}
\endeqn
(ii) Leading terms of
$\Pi_2 (z)$ 
for $z \rightarrow - \infty$: 
\eqn
\Pi_2 (z) = {5 \over 24} - \zeta_3 - {1 \over 4} \ln (-4z)
    + {1 \over z} \biggl [ - {3 \over 4} \ln (-4z) \biggr ]  
    + {\cal O} \left ({1\over z^2} \right ) .
\label{pi21largez}
\endeqn
(iii) The threshold behavior of the imaginary part of $\Pi_2(z)$: 
\eqn
{1 \over \pi} {\rm Im} \Pi_2 (z) 
   = {\pi^2 \over 4} - 2 \delta + {\cal O} (\delta^2)~,
\label{impi21z=1}
\endeqn 
where $\delta = \sqrt{1-1/z}$. This means that the real part of $\Pi_2(z)$
has the logarithmic threshold singularity
of the form
\eqn
\Pi_2 (z) =  -{\pi^2 \over 4} \ln | 1 - z | + \cdots  ~. 
\label{pi21z=1}
\endeqn

To take these analytic properties into account
we construct a function 
$\tilde{\Pi}_2 (z)$ containing five parameters $c_1 , ..., c_5$:
\eqn
\tilde{\Pi}_2 (z) \equiv 
\Pi_2 (z)    
+ (1-z) G(z) ( c_1 + c_2/z ) +c_3 + c_4/z +c_5 F(z) ,
\label{tildepi2}
\endeqn
where $G(z)$ is introduced in (\ref{pi3}) and $F(z) = \ln |1-z|$.
For $z>1$, $F(z)$ has the imaginary part
\eqn
 {\rm Im} F(z) = - \pi ~.
\endeqn 
The coefficients $c_1$, $c_2$ and $c_4$ are determined by requiring 
cancellation of all logarithmic singularities found in  
Eqs. (\ref{pi21largez}) and (\ref{pi21z=1}). 
The remaining $c_3$ and $c_5$ 
are determined by
requiring that $\tilde{\Pi}_2 $ vanishes in the limit $ z \rightarrow 0_-$. 
We find that $\tilde{\Pi}_2 (z)$ is analytic in the $z$-plane
cut along the real axis $(1, \+\infty )$ if we choose 
\eqna
c_1 = {1 \over 2} (1 -\pi^2 ),~~~   
c_2 &=& {1 \over 4} (7 -\pi^2 ),~~~   
c_4 = {1 \over 12} (1 + 5\pi^2 ),~~~   
\nonumber   \\
c_4 &=& {1 \over 4} (-7 +\pi^2 ),~~~   
c_5 = {1 \over 4} \pi^2 .  
\label{coeff}
\endeqna

The function to be Pad\'{e}-approximated is chosen as
\eqn
{1 \over (1+\omega)^2}
\biggl( \tilde{\Pi}_2(z) -\tilde{\Pi}_2(-\infty)\biggr ),
~~~z= {4 \omega \over (1+\omega)^2 }~.
\endeqn
We construct the Pad\'{e} approximant $P(\omega )$ of the form
(\ref{Pomega}) using (\ref{pi21smallz}), (\ref{pi21largez})
and (\ref{impi21z=1}) as the input data. 
We obtained two sets of $a_i$'s and $b_i$'s for each 
[2/3] and [3/2] Pad\'{e} approximations,
which we denote [2/3]A, [2/3]B, [3/2]A, and [3/2]B. 
They are listed in Table II.
The Pad\'{e} results of Sec. 
\ref{subsec:padeapprox}
are obtained from this Table.

As a final check,
we compared directly the imaginary parts of 
the exact  K\"{a}llen-Sabry function and 
its Pad\'{e} approximation. The difference of the two functions
$ \Delta \equiv 
{\rm Im}\Pi_2 ({\rm K-S})/\pi - {\rm Im} \Pi_2 ({\rm Pad\acute{e}})/\pi$
is examined for  the entire momentum range $z \ge 1$.
The fractional deviation $\Delta /({\rm Im}\Pi_2 ({\rm K-S})/\pi)$
is less than 0.2 percent in all cases of the Pad\'{e}
approximations.   The Pad\'{e} approximation functions 
are found to oscillate
around the exact value. 
Also, by integrating $\Delta$ over 
the entire range of $\delta=\sqrt{1-1/z}, ~0 \le \delta \le 1$, we found  
\eqn
\int_0^1 d\delta ~\Delta = \left \{  
\begin{array}{cc}
       ~0.000~276~ &  \cdots {\rm [2/3]A  } \\ 
       -0.000~930~ &  \cdots {\rm [2/3]B  } \\ 
       -0.000~461~ &  \cdots {\rm [3/2]A  } \\ 
       ~0.000~015~ &  \cdots {\rm [3/2]B  }  
\end{array}
\right .
\label{intdel}
\endeqn
Apparently the values of the muon $g-2$ and Lamb shift
are more accurate than are implied by these results.

\begin{table}
\caption{Coefficients of Pad\'{e} approximants for two-loop
vacuum polarization function $\Pi_2(z)$. 
We set $b_0=1$ for the overall normalization. 
\label{table2}
}
\[
\begin{array}{ccccc}
\hline \hline
 {\rm Coefficient}    &   {\rm [2/3]A~ }~~~~   
                      &   {\rm [2/3]B~ }~~~~ 
                      &   {\rm [3/2]A~ }~~~~   
                      &   {\rm [3/2]B~ }~~~~ 
\\ \hline
a_0 &  ~0.218~599~301~    &   ~0.218~599~301~       
    &  ~0.218~599~301~    &   ~0.218~599~301~       
\\ \hline
a_1 &  -0.011~600~062~    &  -0.097~946~773~
    &  -0.051~634~466~    &  ~0.005~017~399~
\\ \hline
a_2 &  -0.118~517~746~    &  -0.176~492~097~ 
    &  -0.190~422~374~    &  -0.143~212~168~
\\ \hline
a_3 &  0                  &   0
    &  -0.023~342~491~    & -0.018~586~700~          
\\ \hline
b_0 &  1                  &   1 
    &  1                  &   1 
\\ \hline
b_1 & -0.228~103~636~     &  -0.623~103~575~ 
    & -0.411~244~220~     &  -0.152~085~728~ 
\\ \hline
b_2 & -0.470~961~931~     &  -0.667~030~155~ 
    & -0.767~838~808~     &  -0.597~234~583~ 
\\ \hline
b_3 &  ~0.060~545~956~     &   ~0.082~509~741~ 
    &  0                  &   0 
\\
\hline \hline
\end{array}
\]
\end{table}

\end{document}